\documentclass[aip,reprint]{revtex4-1}
\usepackage{booktabs}
\usepackage{graphicx}
\usepackage{hyperref}
\draft 

\begin{document}


\title{Reinforcement Learning for Active Matter} 



\author{Wenjie Cai}
\affiliation{School of Physics and Electronics, Hunan University, Changsha 410082, China}

\author{Gongyi Wang}
\affiliation{School of Physics and Electronics, Hunan University, Changsha 410082, China}

\author{Yu Zhang}
\affiliation{School of Physics and Electronics, Hunan University, Changsha 410082, China}

\author{Xiang Qu}
\affiliation{School of Physics and Electronics, Hunan University, Changsha 410082, China}
\affiliation{Department of Physics, King's College London, London WC2R 2LS, United Kingdom}

\author{Zihan Huang}
\email{huangzih@hnu.edu.cn}
\affiliation{School of Physics and Electronics, Hunan University, Changsha 410082, China}



\begin{abstract}
Active matter refers to systems composed of self-propelled entities that consume energy to produce motion, exhibiting complex non-equilibrium dynamics that challenge traditional models. With the rapid advancements in machine learning, reinforcement learning (RL) has emerged as a promising framework for addressing the complexities of active matter. This review systematically introduces the integration of RL for guiding and controlling active matter systems, focusing on two key aspects: optimal motion strategies for individual active particles and the regulation of collective dynamics in active swarms. We discuss the use of RL to optimize the navigation, foraging, and locomotion strategies for individual active particles. In addition, the application of RL in regulating collective behaviors is also examined, emphasizing its role in facilitating the self-organization and goal-directed control of active swarms. This investigation offers valuable insights into how RL can advance the understanding, manipulation, and control of active matter, paving the way for future developments in fields such as biological systems, robotics, and medical science.
\end{abstract}


\maketitle 

\section{Introduction}

Active matter refers to systems in which entities possess intrinsic propulsion mechanisms, enabling them to convert energy into motion.  \cite{gompper20202020,gompper20252025,perez2019bacteria,peng2021imaging,zhang2024interplay,di2011swimming,hu2015modelling,kummel2013circular,di2010bacterial,auschra2021thermotaxis,ma2016reversed,nakano2012molecular,
dreyfus2005microscopic,buttinoni2013dynamical,feinerman2018physics,jadhav2024collective,long2020comprehensive,kondoyanni2022bio,yang2018grand,mcelhenney2024collective,doostmohammadi2018active} These systems, ranging from microscopic self-propelled particles such as bacteria \cite{perez2019bacteria,peng2021imaging,zhang2024interplay,di2011swimming,hu2015modelling,kummel2013circular,di2010bacterial} and colloidal swimmers \cite{auschra2021thermotaxis,ma2016reversed,nakano2012molecular,dreyfus2005microscopic,buttinoni2013dynamical} to larger-scale formations like animal groups \cite{feinerman2018physics,jadhav2024collective}and bio-inspired robotic swarms, \cite{long2020comprehensive,kondoyanni2022bio,yang2018grand} exhibit behaviors that deviate significantly from equilibrium systems due to continuous energy input. Theoretical models of active matter, such as the active Brownian particle (ABP) model \cite{romanczuk2012active,speck2014effective} and continuum theory approaches, \cite{stenhammar2013continuum} have been developed to explain these anomalous behaviors. For instance, an individual active particle exhibits short-time superdiffusion and long-time Fickian diffusion, \cite{howse2007self} which can be captured by introducing a persistent velocity term into the Langevin equation. \cite{datta2024random,huang2015accurate} Similarly, collective behaviors, such as the formation of living crystals \cite{palacci2013living} or motility-induced phase separation, \cite{cates2015motility,buttinoni2013dynamical} can be effectively modeled and described by the ABP model. \cite{omar2021phase} While these models provide a framework for understanding the rich phenomena of active matter, they focus primarily on explaining observed behaviors, rather than providing methods to actively guide or control these dynamics.

Building on the understanding of active matter, the ability to guide or control its dynamics holds significant promise for a range of applications. Controlling the motion of individual active particles allows for precise manipulation in tasks such as autonomous navigation, \cite{alvarez2014computational,jikeli2015sperm} resource search, \cite{abrahms2021emerging,stephens2008decision} and efficient locomotion, \cite{alexander1989optimization,kashiri2018overview} particularly under conditions of uncertainty. \cite{cao2022effective,dabelow2019irreversibility} These capabilities are crucial in micro-robotics, biomedical engineering, and other areas where controlling individual particles can drive advancements in drug delivery, \cite{jain2020overview} nanoscale manufacturing, \cite{zolfaghari2019additive} and environmental sensing. \cite{wang2010nanoparticle} On a larger scale, the regulation of collective behaviors in active matter systems enables the coordination of swarm dynamics for tasks such as collective transport, \cite{tuci2018cooperative} adaptive materials, \cite{lehn2015perspectives} and distributed computing. \cite{mcenroe2022survey} Such control could lead to innovations in fields like robotics, environmental monitoring, and even the development of intelligent materials that adapt to changing external conditions. Therefore, achieving effective control of both individual and collective dynamics in active matter is fundamental for advancing practical applications across various scientific fields.

\begin{figure*}[t]
\centering
\includegraphics[width=0.97\textwidth]{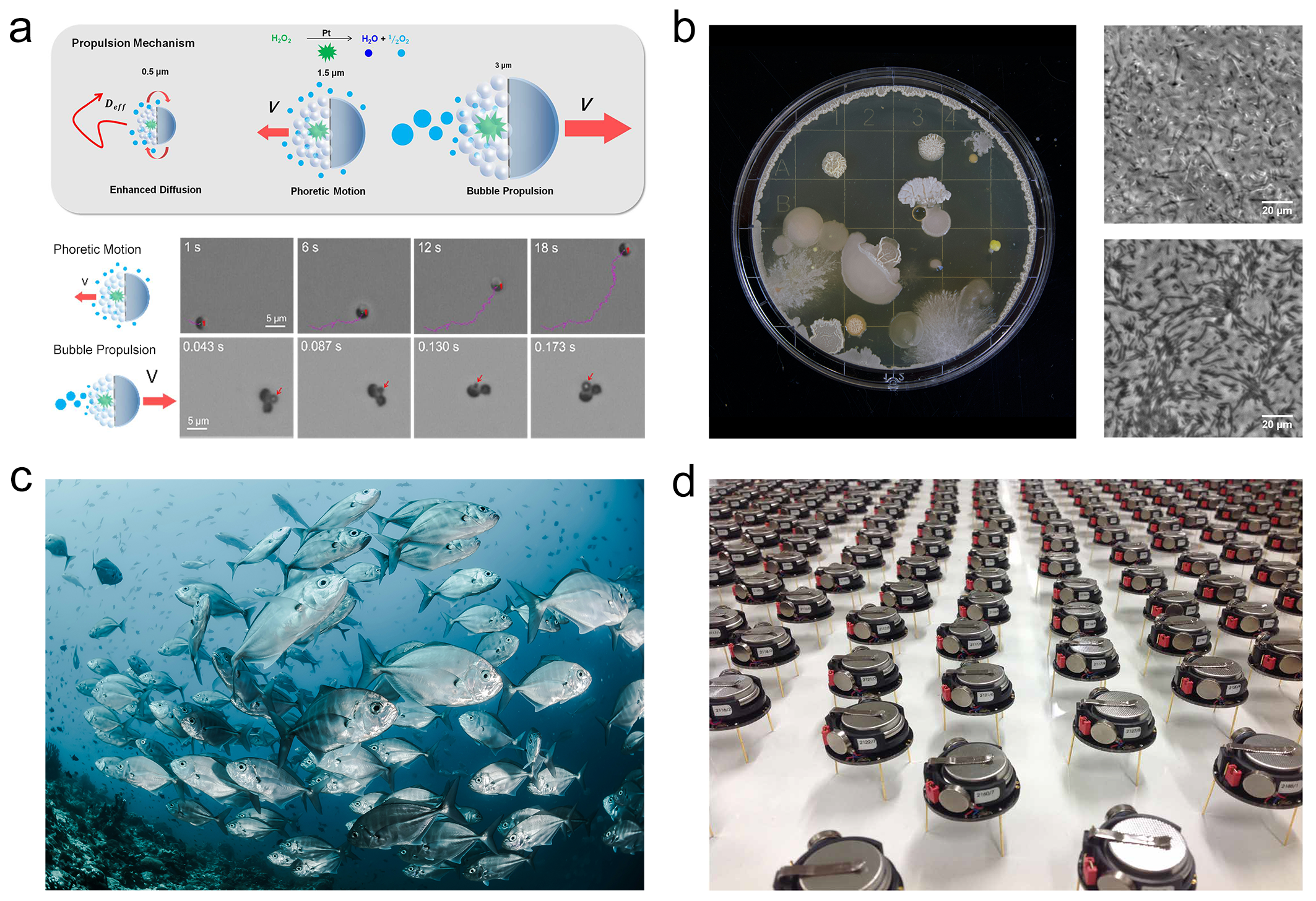}
\caption{Representative types of active matter. (a) Artificial active colloids, shown here as Janus micromotors which are chemically powered, adapted from Ref. \citenum{ma2016reversed}. (b) Microbial systems, depicted with a bacterial suspension (photo by Michael Schiffer on Unsplash) and images of swarming behaviors, adapted from Ref. \citenum{chen2021confinement}. (c) Animals, illustrated with a photo of a school of fish (photo by Sebastian Pena Lambarri on Unsplash). (d) Robotic swarms, adapted from Ref. \citenum{yang2018grand}.}
\label{fig:fig1}
\end{figure*}

Given the challenges posed by the non-equilibrium nature of active matter, reinforcement learning (RL)has emerged as a powerful tool for optimizing and guiding their behaviors. \cite{kaelbling1996reinforcement,li2017deep,yang2024machine,cichos2020machine,nasiri2023optimal} RL offers a robust framework for learning from interactions with the environment, enabling systems to adapt and discover optimal strategies for tasks such as navigation, task allocation, and coordination. Unlike traditional control methods that often rely on predefined models or external inputs, RL facilitates autonomous decision-making by learning policies through trial and error. This capacity for real-time adaptation and optimization makes RL particularly well-suited for the complexities of active matter systems, as it allows them to continuously adjust to dynamic and unpredictable environments. By applying RL to active matter systems, researchers can develop strategies for guiding individual particles through uncertain conditions and for controlling the coordination of large-scale active swarms in collective tasks. Therefore, integrating RL into the study of active matter not only deepens our understanding but also provides a pathway to advancing practical applications that require manipulation and optimization of complex systems. \cite{yang2024machine,cichos2020machine}

In light of these foundations, this review systematically summarizes the integration of RL techniques in the study of active matter. The paper is organized as follows: Section II provides an overview of active matter and RL. Section III examines optimal motion strategies for individual active particles, covering point-to-point navigation problem, foraging strategies, and locomotion strategies, illustrating how RL can optimize movement and decision-making in uncertain environments. Section IV focuses on the regulation of collective dynamics in active swarms, where RL is applied to facilitate the self-organization of active particles and goal-directed control of swarm behaviors. Finally, we conclude the review in Section V, highlighting key insights and suggesting promising directions for future research in this emerging field.

\section{Overview of Active Matter and Reinforcement Learning}

Before addressing the application of RL to active matter systems, we first provide an overview of the key concepts of active matter and RL. Section II.A introduces the foundational aspects of active matter, covering its various types, non-equilibrium behaviors, and propulsion mechanisms. Section II.B then shifts focus to RL, outlining its core concepts and how it enables systems to learn optimal behaviors through interaction with the environment.

\subsection{Active Matter}

Active matter refers to systems in which components are capable of converting energy into motion, driven by intrinsic propulsion mechanisms. These systems span a wide range of scales and types, from microscopic entities like self-propelled colloidal particles and bacteria to larger-scale formations such as animal groups and bio-inspired robotic swarms. Unlike passive systems that tend to reach equilibrium, active matter systems persistently consume energy and exhibit non-equilibrium behaviors such as self-propulsion, motility-induced phase separation, and emergent collective dynamics.

There are various types of active matter, where typical examples are displayed in Fig. 1. Artificial active colloids, such as Janus particles and chemically powered microswimmers [Fig. 1(a)], are typically designed in laboratories and exhibit controlled motion through external fields or chemical reactions. Microbial systems, like bacteria [Fig. 2(b)] or algae, rely on natural propulsion mechanisms such as flagella or cilia, responding to environmental cues for movement. On a larger scale, animal groups, such as schools of fish [Fig. 1(c)] or flocks of birds, demonstrate collective behavior driven by local interactions, resulting in coordinated movement without centralized control. Similarly, bio-inspired robotic swarms [Fig. 1(d)] mimic these natural systems, using a collection of autonomous agents to perform tasks collectively.

The behaviors of active matter are non-equilibrium in nature due to the continuous energy input. \cite{shaebani2020computational,bechinger2016active} Individual active particles often exhibit anomalous diffusion. \cite{cai2025machine,jiang2023switch,qu2024semantic,golestanian2009anomalous} For example, active particles can display superdiffusion at short times, where their displacement scales faster than linearly with time, and Fickian diffusion at long times, where their motion transitions to standard enhanced diffusion. Collective dynamics in active matter systems also demonstrate non-equilibrium behaviors like flock formation, pattern creation, and motility-induced phase separation, where local interactions between particles or agents give rise to global patterns. These non-equilibrium phenomena highlight a key difference from equilibrium systems, where fluctuations are typically averaged out and systems tend to reach a steady state. In contrast, active matter systems exhibit persistent fluctuations and can self-organize into dynamic structures or patterns driven by internal energy consumption.

The study of active matter is essential due to its wide-ranging applications across various fields. For example, active colloids are being explored for use in targeted drug delivery, where their motion can be controlled to navigate complex biological environments. Biological active systems, such as bacteria, hold promise for bio-manufacturing, leveraging their natural propulsion to perform complex tasks. In addition, understanding the collective dynamics of animal groups has applications in developing autonomous systems for robotics and swarm intelligence. Thus, investigating and manipulating the dynamics of active matter is crucial for advancing innovations in diverse fields such as nanotechnology, robotics science, and biomedical engineering.

\subsection{Reinforcement Learning}

\begin{figure}[t]
\centering
\includegraphics[width=0.47\textwidth]{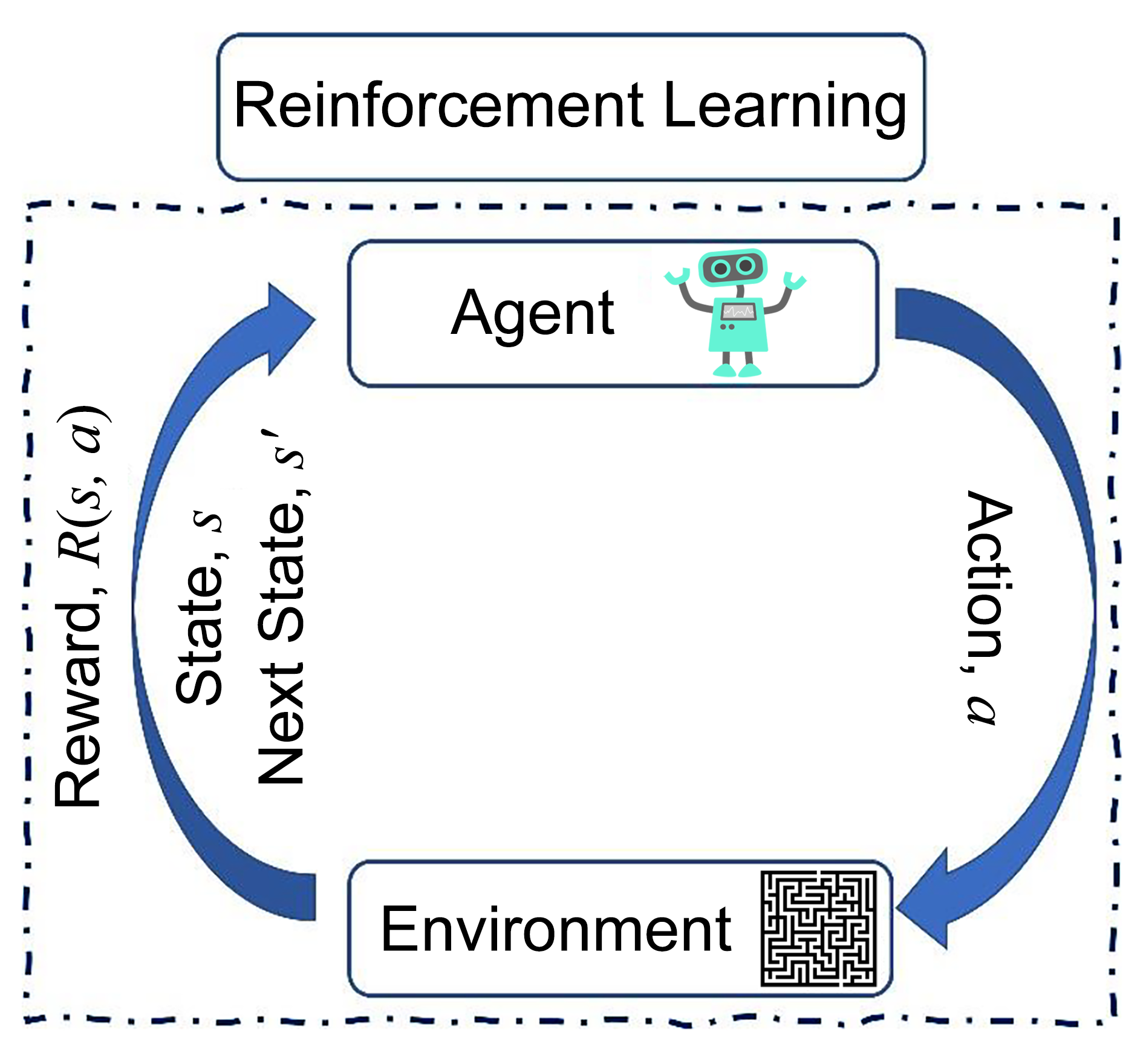}
\caption{Schematic diagram of a standard reinforcement learning system, adapted from Ref. \citenum{jebellat2024reinforcement}.}
\label{fig:fig2}
\end{figure}

\begin{figure*}[t]
\centering
\includegraphics[width=0.95\textwidth]{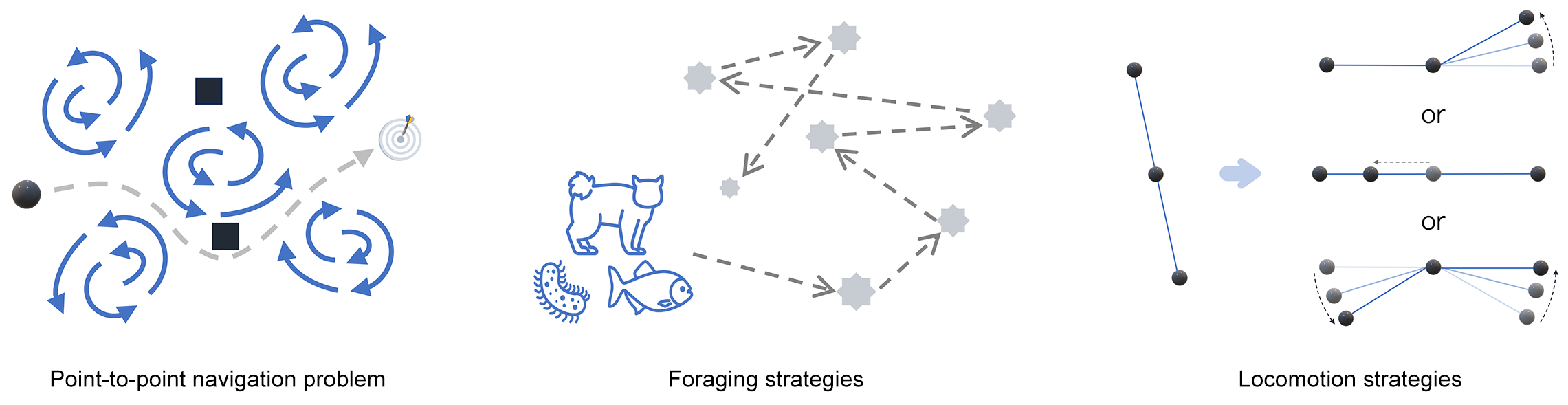}
\caption{Schematic diagram of three aspects regarding the optimal motion strategies for individual active particles. From left to right: point-to-point navigation problem, foraging strategies, and locomotion strategies.}
\label{fig:fig3}
\end{figure*}

Reinforcement learning (RL) is a powerful class of machine learning algorithms that enable agents to learn optimal behaviors through interactions with their environments. \cite{kaelbling1996reinforcement,li2017deep} In RL, an agent takes actions in an environment, receiving feedback in the form of rewards or penalties, and aims to maximize its cumulative reward over time. Unlike supervised learning, where the model is trained on labeled data, RL operates under the paradigm of trial and error, where agents continuously explore different actions to discover the most effective strategy.

RL is typically modeled using a Markov decision process (MDP), \cite{puterman1990markov} which provides a mathematical framework for decision-making problems. As illustrated in Fig. 2, an MDP consists of a set of states $S$, a set of actions $A$, a transition function $P(s'|s, a)$, which defines the probability of moving to state $s'$ after taking action $a$ in state $s$, and a reward function $R(s, a)$, which gives the immediate reward received after taking action $a$ in state $s$. The goal of an RL agent is to learn a policy $\pi(a|s)$, which maps states to actions, in such a way that it maximizes the cumulative reward over time, often measured as the return. This return is typically calculated as the sum of future rewards, often discounted by a factor $\gamma$, representing the agent's preference for immediate rewards over distant ones.

Algorithms of RL can generally be divided into value-based, policy-based, and actor-critic methods. In value-based methods like Q-learning, the agent learns a value function that estimates the expected return for each state-action pair. This value function is used to guide the agent's decisions. Deep Q-network (DQN) \cite{fan2020theoretical} extends Q-learning by utilizing deep neural networks to approximate the Q-value function, enabling the application of RL to complex, high-dimensional state spaces. In policy-based methods, the agent directly learns a policy function that maps states to actions, without needing to learn an explicit value function. One popular policy optimization method is proximal policy optimization (PPO), \cite{schulman2017proximal} which ensures that updates to the policy do not diverge too far from the previous policy, improving stability and sample efficiency. On the other hand, actor-critic methods combine the advantages of both value-based and policy-based approaches. Two models are used here: one for estimating the value function (the critic) and another for estimating the policy (the actor). For example, A2C (advantage actor-critic) \cite{konda1999actor} is a commonly used actor-critic method that improves learning efficiency by considering the advantage function, which measures the relative benefit of taking a specific action in a given state compared to the average behavior.

These RL algorithms provide diverse approaches for learning in complex environments and have been successfully applied to a wide range of scientific tasks. In the following sections, we will discuss the applications of these algorithms in active matter systems.

\section{Optimal motion strategies for individual active particles}

In this section, we explore how RL can be applied to optimize the strategies for motions of individual active particles. As shown in Fig. 3, this includes examining three critical aspects: point-to-point navigation problem, \cite{colabrese2017flow,schneider2019optimal,yang2020micro,alageshan2020machine,muinos2021reinforcement,monderkamp2022active,nasiri2022reinforcement,jiang2023dqn,putzke2023optimal,qiu2024dynamic,amoudruz2024path}
 which focuses on optimal pathfinding in uncertain environments; foraging strategies \cite{colabrese2018smart,morimoto2019foraging,botteghi2021curiosity,wispinski2022adaptive,giammarino2024combining,caraglio2024learning,nasiri2024smart,munoz2024optimal}, where RL is used to enhance the search and collection of resources; and locomotion strategies, \cite{reddy2018glider,tsounis2020deepgait,tsang2020self,liu2021mechanical,zhu2021numerical,zou2022gait,el2023steering,qin2023reinforcement,xue2023exploring,jebellat2024reinforcement,lin2024emergence,flato2024revealing,he2024learning} which aim to optimize how active particles execute movement, including decisions on speed, direction, particle configuration, and other factors, to achieve efficient gait planning and adaptive responses to environmental fluctuations.

\subsection{Point-to-Point Navigation Problem}

The point-to-point navigation problem \cite{li2008point,nasiri2023optimal} involves the task of guiding an agent from an initial position to a target location within an environment, which is subject to various forces such as currents, winds, or other dynamic factors. The goal of this problem is to identify the most efficient path while considering various constraints, such as time, energy consumption, or stability of the system. Traditional solutions to this problem often rely on a variety of well-established techniques, including optimal control theory, \cite{kirk2004optimal} dynamic programming, \cite{zaccone2018ship} and geometrical approaches such as Finsler geometry. \cite{javaloyes2014zermelo} These methods are designed to minimize travel time or energy consumption by providing explicit control strategies that dictate how the agent should move in the environment. However, in active matter systems, these conventional approaches face limitations due to the environmental heterogeneity, random perturbations, and non-equilibrium nature of active particles. This highlights the need for more adaptive approaches that can account for the uncertainties and complexities inherent in active matter systems.

Given the limitations of traditional methods, RL offers a promising solution to the point-to-point navigation problem for individual active particles due to its adaptability to dynamic and uncertain environments. \cite{yang2024machine,cichos2020machine} Unlike traditional methods that rely on predefined models, RL allows active particles to learn optimal navigation strategies through trial and error, adjusting their actions based on real-time feedback. This flexibility enables RL to overcome challenges such as environmental heterogeneity and the non-equilibrium dynamics that characterize active matter systems.

Building upon these expectations, a pioneering study by S. Colabrese et al. investigates the use of RL to optimize the navigation strategies of gravitactic microswimmers in periodic vortex flows, specifically the Taylor-Green vortex flow. \cite{colabrese2017flow} The researchers apply the Q-learning algorithm in numerical experiments to enable the microswimmers to autonomously adjust their swimming direction based on local flow information, allowing them to maximize their vertical displacement. The results demonstrate that these smart microswimmers are able to learn near-optimal navigation strategies through trial and error, effectively escaping flow-induced trapping regions and exploiting the ``fluid elevator" effect for more efficient ascent. This RL-based approach significantly outperforms traditional passive gravitactic strategies, which often fail to overcome the challenges of strong vortex regions and shear flow zones. Additionally, the study shows that the microswimmers exhibit adaptability to variations in the flow field, highlighting the versatility of RL in optimizing navigation strategies under changing environments.

\begin{figure*}[t]
\centering
\includegraphics[width=0.95\textwidth]{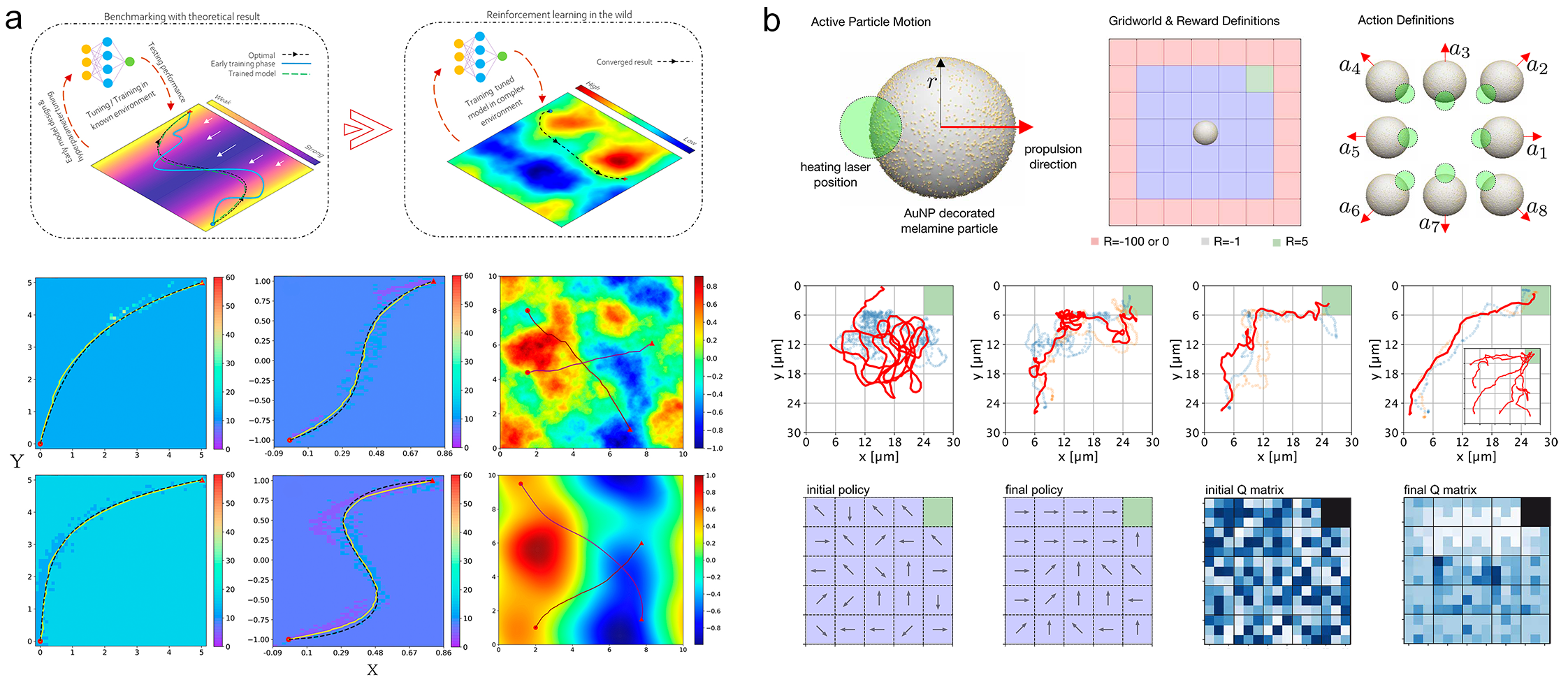}
\caption{Representative examples of RL-based point-to-point navigation for individual active particles. (a) A policy-gradient RL approach identifies near-optimal trajectories, adapted from Ref. \citenum{nasiri2022reinforcement}. Top: schematic illustrating the RL scheme; bottom: sample learned paths in various flow fields. (b) An experimental RL approach controls gold nanoparticle-coated microswimmers propelled by self-thermophoresis, adapted from Ref. ~\citenum{muinos2021reinforcement}. Top: schematic of the microswimmer and a gridworld-based RL design; bottom: trajectories showing how Q-learning converges to efficient navigation strategies under real-world noise and delays.}
\label{fig:fig4}
\end{figure*}

After that, M. Nasiri et al. introduce a deep RL approach using the A2C algorithm, for guiding active particles toward a target in two-dimensional force and flow fields in simulations, \cite{nasiri2022reinforcement} as shown in Fig. 4(a). By discretizing the environment into a gridworld representation, their method avoids the need for reward shaping, allowing the active particle to asymptotically learn near-optimal paths purely from experience and to replicate known analytical solutions for simpler setups. Moreover, it successfully handles more challenging scenarios, including Gaussian random potentials in which classical path-planning algorithms often fail. In another representative simulation work, M. Putzke et al. employ a tabular Q-learning algorithm that relies solely on the distance and bearing to the target, demonstrating time-optimal navigation of active particles through potential barriers, uniform flows, Poiseuille flows, and swirling flows. \cite{putzke2023optimal} Notably, their results remain robust to moderate orientational noise, indicating that Q-learning can maintain high performance under realistic conditions.

Following these advancements, S. Mui\~{n}os-Landin et al. extend RL applications beyond simulations by implementing them in real experimental conditions. \cite{muinos2021reinforcement} As illustrated in Fig. 4(b), their study demonstrates the feasibility of RL-based navigation control for artificial gold nanoparticle-coated microswimmers propelled via self-thermophoresis in an aqueous environment . Unlike simulated agents, these microswimmers face significant challenges, including Brownian motion, feedback delays, and external noise. To address these issues, the researchers design a discrete gridworld framework and employ a Q-learning algorithm to guide the microswimmers toward a designated target through laser-induced propulsion. Despite the stochasticity of their microscopic surroundings, the microswimmers successfully learn optimal navigation strategies purely from interaction with the environment, showing the robustness of RL in real-world active matter systems. This experimental validation marks an important step toward integrating reinforcement learning with active particle control at the microscale in reality, bridging the gap between theoretical models and physical implementations.

In addition to these studies highlighted above, a range of other RL-based works on the point-to-point navigation problem for individual active particles can be found in Table I. Taken together, these works broaden the scope of RL-based navigation in active matter systems, underscoring RL's capability to deal with the complexity of non-equilibrium environments. Further refinements of existing algorithms and deeper integration of experimental feedback are likely to advance the field even more. Potential directions include designing adaptive reward structures to handle multi-objective tasks (e.g., balancing speed and energy efficiency), and expanding to multi-agent systems in which large numbers of active particles coordinate to achieve shared goals. These developments will be essential for realizing robust, real-world applications of RL-driven navigation at both microscopic and macroscopic scales.

\begin{figure*}[t]
\centering
\includegraphics[width=0.95\textwidth]{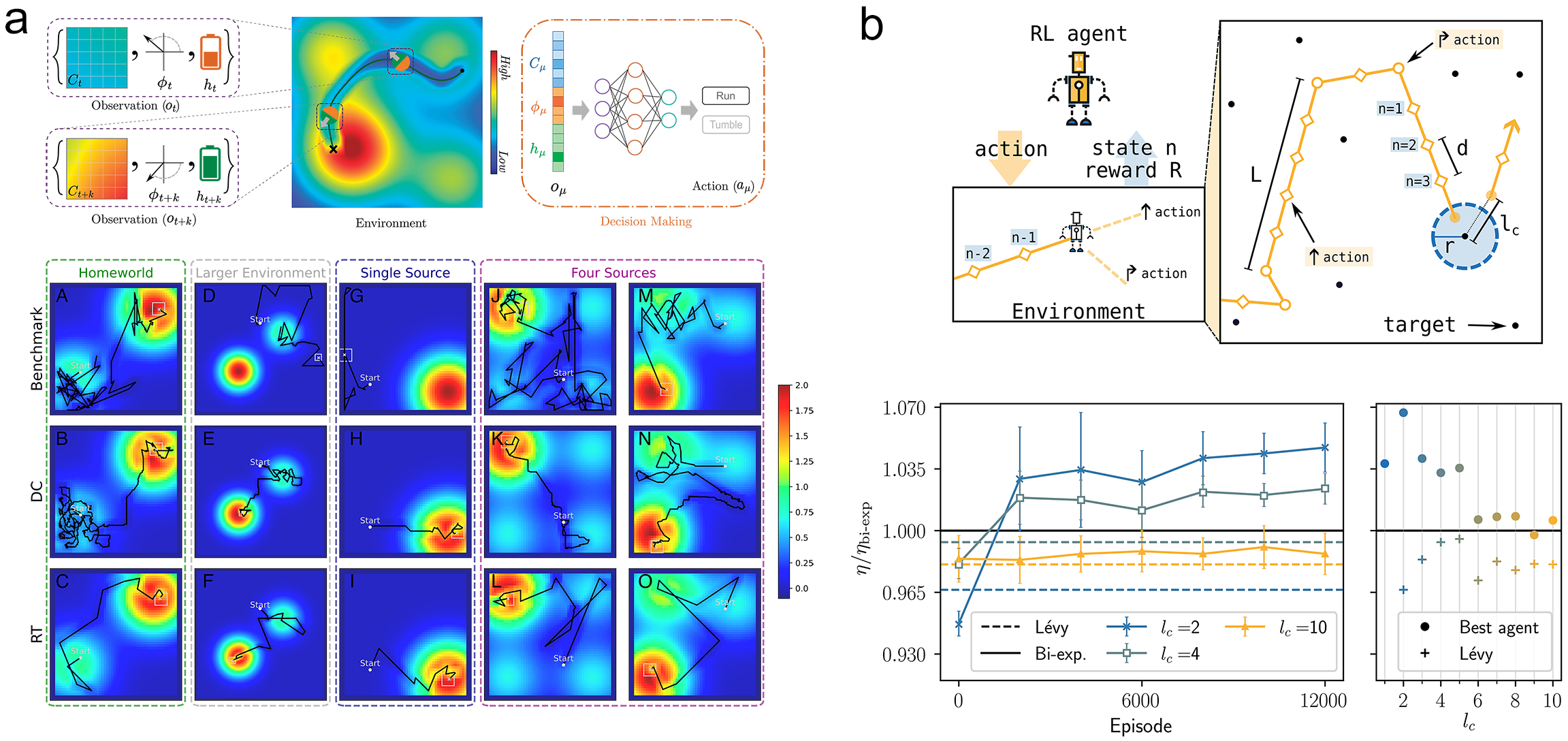}
\caption{RL-based foraging strategies in active matter systems. (a) A deep Q-learning approach optimizes nutrient foraging strategies for smart active particles, adapted from Ref. \citenum{nasiri2024smart}. Top: Schematic diagram of the RL-based approach. Bottom: Example trajectories of the benchmark model, DC agent, and RT agent in different environments. (b) A projective simulation-based RL approach models non-destructive foraging behaviors in randomly distributed target environments, adapted from Ref. \citenum{munoz2024optimal}. Top: Schematic representation of the RL framework. Bottom: Learning curves comparing the efficiency of RL agents and benchmark strategies.}
\label{fig:fig5}
\end{figure*}

\subsection{Foraging Strategies}

Foraging strategies refer to the set of behaviors and decision-making processes employed by organisms to locate and acquire resources from their environment. Unlike the point-to-point navigation problem, which typically involves traveling from a known starting point to a target destination, foraging requires individuals to seek out and gather resources that are often dispersed, variable, and uncertain. Foraging behaviors in nature can be observed across various scales, from microorganisms searching for nutrients to animals hunting for food in vast and complex environments.

Traditional methods for solving foraging problems often rely on established models such as L\'{e}vy walks and Brownian motion-based approaches, \cite{jandhyala2018applications} which attempt to mimic the efficient foraging patterns observed in biological organisms. These methods typically assume static environments, and their performance depends on predefined rules for movement, often based on simplistic assumptions about resource distributions. However, in real-world settings, resources may be scarce, transient, or dynamically changing. This highlights the need for more flexible, adaptive foraging strategies.

In particular, RL fundamentally addresses sequential decision-making, where an agent incrementally refines its behavior through trial-and-error learning from feedback. This process mirrors the adaptive strategies employed by biological organisms as they enhance their foraging efficiency. In natural systems, foragers balance exploration (seeking new resources) and exploitation (utilizing known resources), a trade-off that aligns closely with the core principle of RL. As the environment changes dynamically, RL offers a significant advantage over traditional methods by continuously adjusting strategies based on real-time environmental feedback.

At the microscopic scale, several studies have applied RL to optimize foraging strategies for active particles. As shown in Fig. 5(a), M. Nasiri et al. employ deep Q-learning to train smart active particles for efficient nutrient foraging within a simulated environment. \cite{nasiri2024smart} The study considers agents with limited sensory input, specifically, they can only perceive local nutrient concentrations, their health status, and their current orientation, without global knowledge of the environment. The authors compare two control models: the run-and-tumble (RT) model and the direction-controlling (DC) model. The results show that the RL-trained agents outperform traditional random search strategies, including L\'{e}vy walks and chemotaxis, by learning to exploit correlations within the environment that are initially unknown. Moreover, the trained agents exhibit strong generalization capabilities, successfully applying their foraging behaviors to unfamiliar, structurally different environments. Another work is conducted by M. Caraglio et al., \cite{caraglio2024learning} which focuses on optimizing the target-search strategies of intermittent ABPs with RL algorithms. In this study, the ABPs are capable of switching between two distinct modes: a passive Brownian motion mode and an active Brownian motion mode. Projective simulation is employed to learn efficient switching policies, in which particles decide how long they stay in either the passive or active phase primarily according to how long they have already been in the current phase and whether they have found the target. The researchers find that the target-search efficiency increases with the self-propulsion of the particles during the active phase. Interestingly, the optimal duration of the passive phase decreases monotonically with increased activity, while the optimal duration of the active phase exhibits non-monotonic behavior, peaking at intermediate P\'{e}clet numbers.

On the other hand, for animal foraging at the macroscopic scale, RL has also shown substantial potential for modeling adaptive search behaviors. For instance, G. Mu\~{n}oz-Gil et al. develop an RL framework to explore optimal animal foraging strategies by modeling agents that learn to forage in environments with randomly distributed targets, \cite{munoz2024optimal} as illustrated in Fig. 5(b). The study focuses on a non-destructive foraging model where the agent must search for replenishable targets. The model employs the projective simulation algorithm to maximize the search efficiency, with the agent choosing at each step to either continue in the same direction or turn to a new random direction. Numerical experiments demonstrate that the RL agents learn strategies that surpass the efficiency of known models like L\'{e}vy walks and bi-exponential distributions [bottom of Fig. 5(b)], providing valuable insight into the learning process of biological agents. Similarly, N. J. Wispinski et al. apply deep RL to study patch foraging, \cite{wispinski2022adaptive} a fundamental optimization problem in ecology and animal foraging. In this work, agents learn to adaptively adjust their patch-leaving behaviors via RL in a continuous 3D environment. Their learned policies are then compared against the marginal value theorem (MVT), a well-known theoretical solution for patch foraging, to assess how close they come to optimal behavior. Maximum a posteriori policy optimization algorithm is used here to train agents in a continuous 3D foraging environment. Agents are tasked with deciding when to leave a patch and travel to a new one based on the decaying rewards within each patch. The results show that the agents adapt their foraging strategy according to the distance between patches, aligning with biological foragers' behavior. When accounting for temporal discounting, the trained agents approach optimal foraging behavior as predicted by the MVT.

In addition, a growing body of RL-based research on foraging strategies for active particles is available in Table I. These works collectively highlight RL's ability to learn adaptive search strategies in complex, dynamically changing environments. Importantly, future research will benefit from enhanced experimental validations, bridging the gap between numerical results and real-world observations. This will involve testing RL-driven foraging strategies in actual environments, refining algorithms based on experimental feedback, and validating their generalizability across diverse biological systems.

\subsection{Locomotion Strategies}

Locomotion strategies refer to the modes of active particles to execute movement, including decisions on speed, direction, particle configuration, and other factors, with the goal of achieving efficient gait planning and adaptive responses to environmental fluctuations. \cite{zou2022gait,reddy2018glider} Unlike foraging or navigation, which are concerned with reaching specific targets or acquiring resources, locomotion focuses on optimizing how particles move in their environment. This involves control over both internal dynamics and external interactions, enabling particles to adapt their movement in response to changing conditions. The study of locomotion strategies is critical for designing artificial systems with effective, adaptive behaviors that can function in real-world, dynamic environments. Understanding and engineering efficient locomotion patterns is particularly important in applications ranging from microscopic robots to autonomous aerial vehicles, as it can lead to enhanced performance, energy efficiency, and robustness.

\begin{figure*}[t]
\centering
\includegraphics[width=0.95\textwidth]{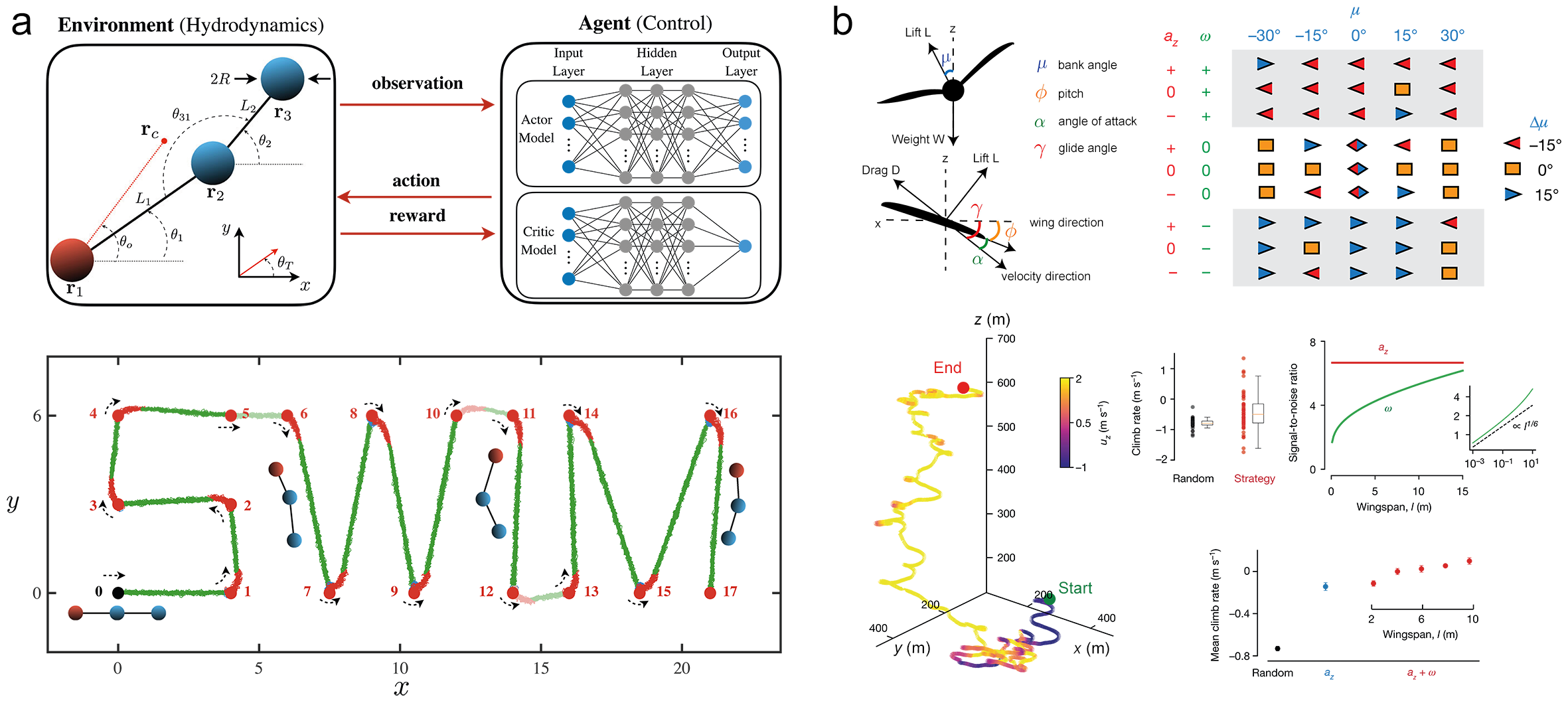}
\caption{RL-based locomotion strategies for active matter. (a) A deep RL approach for controlling a microswimmer with rod-sphere structures, adapted from Ref. \citenum{zou2022gait}. Top: Schematic of the swimmer's configuration and deep neural network architecture. Bottom: Demonstration of the microswimmer autonomously navigating a complex trajectory using various locomotion gaits. (b) A RL-based soaring strategy for a glider navigating atmospheric thermals, adapted from Ref. \citenum{reddy2018glider}. Top: Force-body diagram of the glider and the learned control policy. Bottom: Experimental results showing the glider's performance with the learned locomotion strategy, including trajectory, climb rate comparison with random control, and the effect of wingspan on performance.}.
\label{fig:fig6}
\end{figure*}

\begin{turnpage}
\begin{table*}[htbp]
\centering
\setlength{\tabcolsep}{5pt}
\scriptsize

\caption{Overview of representative RL studies regarding optimal motion strategies for individual active particles.}
\label{tab:study_targets}
\begin{ruledtabular}
\begin{tabular*}{\textheight}{@{\extracolsep\fill}p{8.85cm}p{1.5cm}p{7.7cm}p{0.95cm}p{2.0cm}p{0.7cm}}
\textbf{Research Subject} & \textbf{Category} & \textbf{Objective} & \textbf{Method} & \textbf{RL Algorithm} & \textbf{Year}\\

        Smart microswimmers in Taylor-Green vortex flows \cite{colabrese2017flow}
        & Navigation
        & Maximize long-term vertical displacement
        & Sim.
        & Q-learning
        & 2017 \\

        Smart active particles in potential landscapes \cite{schneider2019optimal}
        & Navigation
        & Minimize travel time
        & Sim.
        & Q-learning
        & 2019 \\

        Brownian micro/nano motor \cite{yang2020micro}
        & Navigation
        & Minimize travel time and localization errors
        & Sim.
        & DDPG
        & 2020 \\

        Microswimmers in turbulent flows \cite{alageshan2020machine}
        & Navigation
        & Minimize travel time
        & Sim.
        & Q-learning
        & 2020 \\

        Self-thermophoretic artificial microswimmers \cite{muinos2021reinforcement}
        & Navigation
        & Minimize travel time
        & Exp.
        & Q-learning
        & 2021 \\

        Self-propelled active particles in complex environments \cite{nasiri2022reinforcement}
        & Navigation
        & Minimize travel time
        & Sim.
        & A2C
        & 2022  \\

        Active Brownian particles in complex motility fields \cite{monderkamp2022active}
        & Navigation
        & Minimize travel time while avoiding low motility zones
        & Sim.
        & Q-learning
        & 2022 \\

        Homogeneous microrobots of varying scales \cite{jiang2023dqn}
        & Navigation
        & Minimize travel time while avoiding obstacles
        & Exp.
        & DQN
        & 2023 \\

        Smart active particle in complex flow fields and potential landscapes \cite{putzke2023optimal}
        & Navigation
        & Minimize travel time
        & Sim.
        & Q-learning
        & 2023 \\

        Magnetic helical microrobots \cite{qiu2024dynamic}
        & Navigation
        & Minimize travel time while avoiding obstacles and regulating velocity
        & Exp.
        & PPO
        & 2024 \\

        Artificial bacterial flagella in retinal capillaries \cite{amoudruz2024path}
        & Navigation
        & Minimize travel time and trajectory length
        & Sim.
        & V-racer
        & 2024 \\


        Inertial particles in vortical flows \cite{colabrese2018smart}
        & Foraging
        & Maximize particle's local vorticity
        & Sim.
        & Q-learning
        & 2018 \\

        Tephritid fruit fly larvae under nutrient variability \cite{morimoto2019foraging}
        & Foraging
        & Maximize nutrient acquisition
        & Sim.
        & UCB \& TS
        & 2019 \\

        Curiosity-driven RL agent \cite{botteghi2021curiosity}
        & Foraging
        & Maximize exploration efficiency and map completeness
        & Sim.
        & DDPG
        & 2021 \\

        Human foraging behavior data in virtual open-field environments \cite{wispinski2022adaptive}
        & Foraging
        & Maximize search efficiency
        & Exp.
        & PPO
        & 2022 \\

        Biological foragers in ecological patch environments \cite{giammarino2024combining}
        & Foraging
        & Maximize cumulative temporally discounted rewards
        & Sim.
        & PPO
        & 2023 \\

        Intermittent active Brownian particles in homogeneous environments \cite{caraglio2024learning}
        & Foraging
        & Minimize target search time
        & Exp.
        & PS
        & 2024 \\

        Animal in sparsely distributed immobile target environments \cite{munoz2024optimal}
        & Foraging
        & Maximize search efficiency
        & Sim.
        & PS
        & 2024 \\

        Smart active particles in unknown environments \cite{nasiri2024smart}
        & Foraging
        & Maximize nutrient collection efficiency
        & Sim.
        & Q-learning
        & 2024 \\


        Autonomous glider with a 2-meter wingspan \cite{reddy2018glider}
        & Locomotion
        & Maximize vertical wind acceleration
        & Exp.
        & Q-learning
        & 2018 \\

        Quadrupedal robots over rigid, non-flat terrain \cite{tsounis2020deepgait}
        & Locomotion
        & Minimize foothold deviation, orientation misalignment, and locomotion effort
        & Sim.
        & PPO
        & 2020 \\

        Rod-sphere structure microswimmers \cite{tsang2020self}
        & Locomotion
        & Maximize the swimmer's cumulative displacement scaled by its radius
        & Sim.
        & Q-learning
        & 2020\\

        Rod-sphere structure microswimmers \cite{liu2021mechanical}
        & Locomotion
        & Maximize net angular displacement of machine's centroid
        & Sim.
        & Q-learning
        & 2021\\

        Fish-like robotic swimmer \cite{zhu2021numerical}
        & Locomotion
        & Minimize the distance to the target
        & Sim.
        & DQN
        & 2021 \\

        Rod-sphere structure microswimmers \cite{zou2022gait}
        & Locomotion
        & Maximize displacement along the target direction while penalizing directional deviations
        & Sim.
        & PPO
        & 2022\\

        Thin and deformable active filaments in a fluid \cite{el2023steering}
        & Locomotion
        & Minimize required time or energy
        & Sim.
        & Q-learning
        & 2023 \\

        Multi-link swimmers \cite{qin2023reinforcement}
        & Locomotion
        & Maximize horizontal displacement of the swimmer's centroid per cycle while minimizing directional deviations
        & Sim.
        & Q-learning
        & 2023\\

        Storm petrels \cite{xue2023exploring}
        & Locomotion
        & Maximize performance while minimize cost
        & Sim.
        & DDPG
        & 2023\\

        Rod-sphere structure microswimmers \cite{jebellat2024reinforcement}
        & Locomotion
        & Maximize net mechanical displacement of microrobots per propulsion cycle
        & Sim.
        & Q-learning
        & 2024 \\

        Elastic three-sphere microswimmer immersed in a viscous fluid \cite{lin2024emergence}
        & Locomotion
        & Maximize forward displacement
        & Sim.
        & DQN
        & 2024 \\

        Autonomous thermal soaring under horizontal wind conditions \cite{flato2024revealing}
        & Locomotion
        & Maximize altitude gain while minimizing distance to the thermal center
        & Sim.
        & DDPG
        & 2024\\

        Eight-legged robots on rugged terrains \cite{he2024learning}
        & Locomotion
        & Maximize forward speed while minimizing lateral displacement
        & Exp.
        & PPO
        & 2024\\

\end{tabular*}
\end{ruledtabular}
\footnote{Meanings of abbreviations in the table are listed as follows. RL: Reinforcement learning. Sim.: Simulation; Exp.: Experiment; DDPG: Deep deterministic policy gradient; A2C: Advantage actor-critic; DQN: Deep Q-network; PPO: Proximal policy optimization; UCB: Upper confidence bound; TS: Thompson sampling; PS: Projective simulation.}
\end{table*}
\end{turnpage}

Traditional approaches to finding the optimal locomotion strategy primarily rely on physics-based models that describe particle movement through mechanical equations and kinematic analysis. For example, using Newtonian mechanics or Lagrangian dynamics, \cite{toschi2009lagrangian} researchers can simulate the movement of particles in various environments. \cite{toschi2009lagrangian} While these methods can provide accurate predictions for simple systems, they encounter limitations in more complex, dynamic environments. In practical scenarios, particles may experience non-homogeneous flow fields, fluid dynamics effects, or unpredictable environmental changes, making traditional motion models ineffective in addressing these complexities. Moreover, these methods typically rely on predefined movement models or control strategies, which are constrained by the accuracy and adaptability of the underlying assumptions. When faced with dynamic and uncertain environments, traditional methods struggle to provide real-time, adaptive movement decisions. In contrast, as RL does not require precise physical models but instead optimizes behavior through agent-environment interactions, it offers a promising alternative for studying locomotion strategies, particularly in situations where traditional methods fail to account for dynamic and stochastic factors.

Locomotion strategies for active matter encompass various categories, including active particles with adjustable configurations and soaring flight dynamics. For the former, the most representative example is the microswimmer with rod-sphere structures [top of Fig. 6(a)], an active system consisting of rods and spherical segments connected to allow net motion. By dynamically adjusting parameters such as joint angles and coupling stiffness during motion, these microswimmers can switch between different locomotion modes, enabling agile navigation and robust control in complex environments. For instance, in the work by Z. Zou et al., \cite{zou2022gait} a deep RL approach is employed to enable a microswimmer with rod-sphere structures to autonomously learn effective locomotory gaits for navigation in low-Reynolds-number environments. By training the system using the PPO algorithm, the microswimmer can switch between translation, rotation, and combined modes without relying on predefined gait patterns. Notably, the RL-based locomotion strategies allow the microswimmer to follow designated paths and perform target navigation [bottom of Fig. 6(a)]. This robust path-tracking ability highlights the versatility of the microswimmer and its potential for applications in biomedical fields such as targeted drug delivery and microsurgery.

Similarly, Y. Liu et al. explore RL to optimize the rotational motion of a rod-sphere microswimmer with a three-sphere chain configuration. \cite{liu2021mechanical} Their study demonstrates that, regardless of the number of spheres, RL converges to a ``traveling wave policy" that governs the rotation of the microswimmer. This locomotion strategy allows the system to perform efficient rotational movements, with RL dynamically adjusting the sequence of sphere actuations to ensure optimal rotation. The results show that even as the number of spheres increases, the system can still develop effective locomotion strategies through RL. In another work, L. Lin et al. employ the DQN to optimize the locomotion of an elastic three-sphere microswimmer. \cite{lin2024emergence} This investigation focuses on developing an effective motion strategy for the microswimmer by utilizing RL to control the spring dynamics between the spheres. The RL-based system autonomously learns a ``waiting strategy", where the microswimmer pauses at certain points in its motion, allowing the springs to relax before continuing the motion. This strategy is particularly important at high actuation velocities, where the system otherwise struggles with performance degradation.

On the other hand, soaring flight dynamics represents another important category of locomotion strategies for active matter. Soaring dynamics, primarily observed in gliders and birds, focuses on exploiting natural atmospheric phenomena, such as thermals and updrafts, to sustain flight with minimal energy. This approach involves adaptive flight control, where systems learn to navigate fluctuating environmental conditions, like wind or thermal gradients, to optimize movement. Recent studies, particularly using RL, have advanced the development of autonomous aerial systems capable of efficiently exploiting these dynamics.

As displayed in Fig. 6(b), a representative study by G. Reddy et al. trains an autonomous glider to navigate atmospheric thermals using the Q-learning algorithm. \cite{reddy2018glider} The glider, equipped with a flight control system that adjusts bank angle and pitch, learns to optimize its flight path in turbulent environments. By leveraging feedback from the environment, the system autonomously improves its climb rate compared to random strategies. The study emphasizes the use of vertical wind accelerations and roll-wise torques as mechanosensory cues, which guide the glider's movement and enable it to efficiently exploit thermals for sustained flight. Through repeated trials, the glider autonomously improves its detection of thermal updrafts and its navigation strategy to stay within them, demonstrating how RL can enable optimal flight strategies in dynamic, real-world conditions. After that, Y. Flato et al. employ deep RL to investigate thermal soaring under horizontal wind conditions. \cite{flato2024revealing} Using the deep deterministic policy gradient algorithm, they enable a glider to autonomously learn how to locate and remain within thermal updrafts. The study identifies two key learning challenges: achieving stable flight and staying close to the thermal center. To overcome these, reward shaping is used to gradually introduce more complex wind conditions during training. The learned policy mirrors the soaring strategies of real-world vultures, showing how RL can replicate and optimize animal-like flight behaviors in complex environments.

Beyond the works discussed, several other RL-based studies on locomotion strategies for active matter, including those focused on multi-legged systems \cite{he2024learning,tsounis2020deepgait} and bio-inspired robotics, \cite{xue2023exploring} are summarized in Table I. Future research could advance the field by developing RL-based locomotion strategies that enhance robustness to adversarial environments, ensuring that systems maintain efficient movement even in unpredictable or adverse conditions. Additionally, learning under uncertainty is a critical area for improving locomotion strategies, as agents often need to move through noisy, incomplete, or changing environmental data. Overcoming these challenges will be pivotal in expanding the application of RL-driven locomotion strategies, with promising implications for autonomous systems in fields like environmental monitoring and space exploration.

\section{Regulation of collective dynamics in active swarms}

In this section, we examine how RL can regulate and control the collective dynamics in active swarms, focusing on two complementary aspects. First, we discuss the self-organization of active swarms, \cite{loffler2023collective,grauer2024optimizing,costa2020automated,wang2023modeling,durve2020learning,nuzhin2021animals,verma2018efficient,yu2021swarm,sunehag2019reinforcement,pinsler2018inverse,hahn2019emergent} where RL helps individual behaviors optimize local interactions, leading to the emergence of complex patterns like flocking or clustering, without direct centralized control or external influence. Second, we explore the goal-directed control of swarm behaviors, \cite{falk2021learning,schrage2023ultrasound,heuthe2024counterfactual,shen2022deep} where RL facilitates adjustments to global intervention parameters, guiding individual agents to align with predefined collective goals through external influence or manipulation.

\subsection{Self-Organization}

Self-organization in active swarms refers to the spontaneous emergence of ordered collective behaviors from the local interactions between active particles, without the need for centralized control or direct external influence. \cite{karsenti2008self,dombrowski2004self,andersen2009self,yu2022deep} These behaviors, such as flocking, clustering, or pattern formation, arise through decentralized decision-making processes based on local information exchanges between agents. In natural systems, such self-organizing phenomena are commonly observed in biological systems, such as schools of fish, \cite{niwa1994self} flocks of birds, \cite{papadopoulou2022self} or insect colonies, where simple individual rules can lead to highly complex and coordinated group behavior.

\begin{figure*}[t]
\centering
\includegraphics[width=0.95\textwidth]{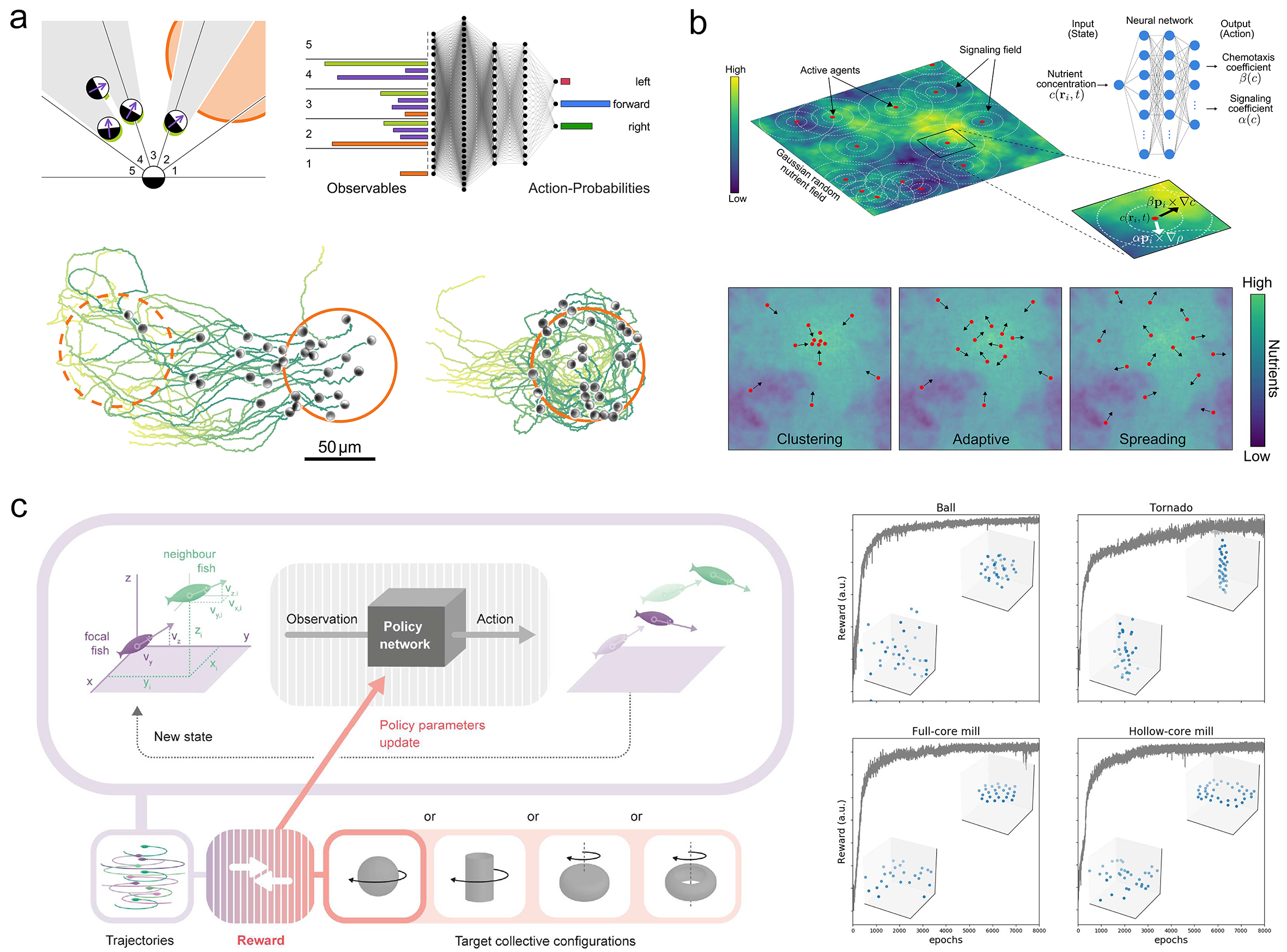}
\caption{Self-organization driven by RL in active swarms. (a) Top: Light-responsive active colloidal particles using RL to optimize foraging behaviors. Bottom: Experimental snapshots of trained particles demonstrating self-organized collective behaviors like flocking and milling during food foraging. Figures are adapted from Ref. \citenum{loffler2023collective}. (b) Top: Schematic of communicating active particles in a nutrient field, utilizing RL to optimize movement based on local nutrient and signaling molecule concentrations. Bottom: Illustration of the three learned collective strategies: clustering, adaptive, and spreading. Figures are adapted from Ref. \citenum{grauer2024optimizing}. (c) Left: The RL framework for fish schooling, where a policy network determines motion based on sensory inputs. Bottom: Training results showing the emergence of four distinct collective motion patterns: rotating ball, tornado, full-core mill, and hollow-core mill. Figures are adapted from Ref. \citenum{costa2020automated}.}
\label{fig:fig7}
\end{figure*}

\begin{figure*}[t]
\centering
\includegraphics[width=0.95\textwidth]{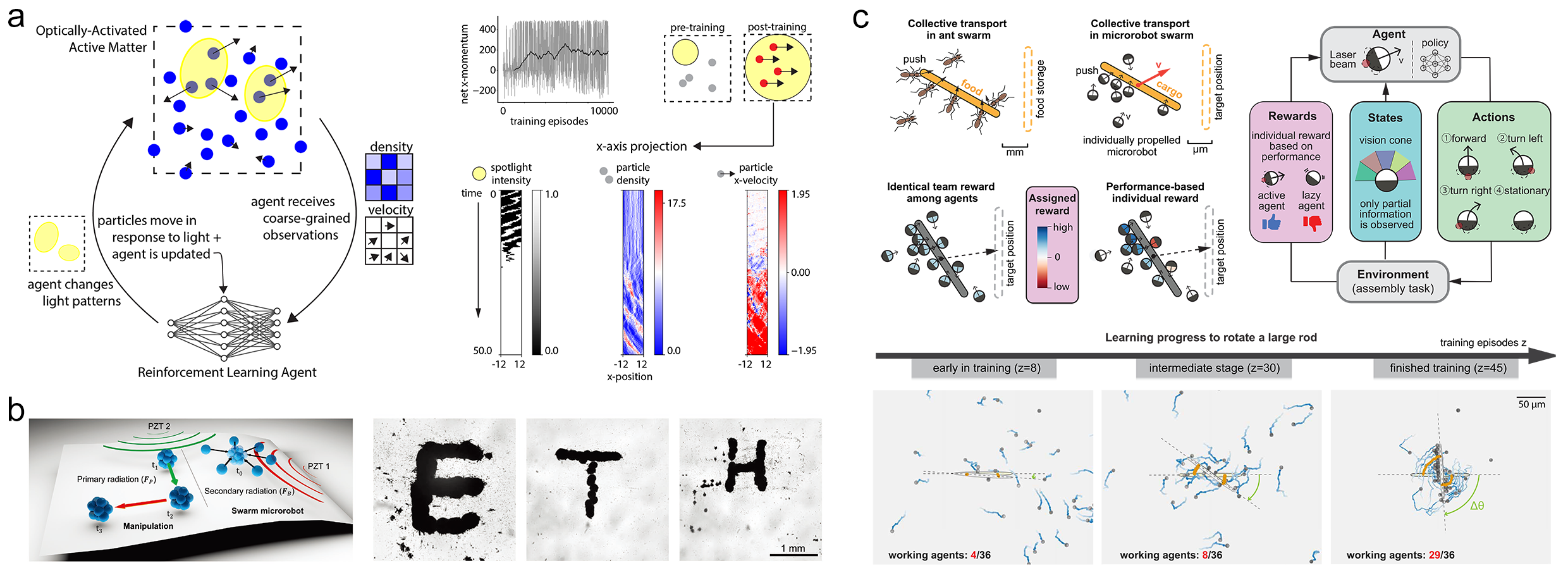}
\caption{Goal-directed control of active swarms driven by RL. (a) Left: Schematic of the RL framework for controlling activity in self-propelled disks through localized light modulation. Bottom: Experimental results showing directional transport in self-propelled disks. Figures are adapted from Ref. \citenum{falk2021learning}. (b) Left: Illustration of the swarm microrobots formed by microbubbles and manipulated by primary and secondary radiation forces. Bottom: Swarm microrobots spelling ``ETH" through ultrasound-guided motion. Figures are adapted from Ref. \citenum{schrage2023ultrasound}. (c) Top: An RL approach for controlling microrobots to perform collective transport. Bottom: Experimental snapshots of the swarm's behavior during training as they collectively rotate and transport a rod. Figures are adapted from Ref. \citenum{heuthe2024counterfactual}.}
\label{fig:fig8}
\end{figure*}

Traditionally, the study of self-organization in active matter systems has relied on physics-based models, such as the Vicsek model, \cite{vicsek1995novel} which describes the alignment of agents based on local interactions and neighborhood rules. These models capture the essence of self-organized behaviors through simplified assumptions, such as alignment or repulsion among agents, and have provided important insights into the conditions under which such behaviors emerge. However, traditional approaches often face significant challenges when it comes to capturing the full complexity of real-world systems. These models are typically limited by the assumptions of fixed or simple interaction rules, making it difficult to account for the dynamic and adaptive nature of real-world active matter. Additionally, these models may not generalize well to systems with complex or changing environments, where interactions between agents are more diverse and require more flexible strategies. Particularly, RL has become a widely used approach for studying self-organization in active swarms, as it does not rely on predefined interaction rules. By allowing active particles to autonomously adjust their motion behaviors based on local interactions, RL facilitates the exploration of emergent phenomena in more realistic and complex environments.

Recent studies have explored the role of RL in modeling self-organization in both microscopic and macroscopic active matter systems. At the microscopic scale, investigations such as those by R. L\"{o}ffer et al. \cite{loffler2023collective} and J. Grauer et al. \cite{grauer2024optimizing} leverage RL to optimize motion behaviors of active particles and discover emergent collective patterns. As shown in Fig. 7(a), L\"{o}ffer et al. focus on light-responsive active colloidal particles, applying the PPO algorithm to optimize individual foraging strategies based on limited sensory inputs, such as a 180$^\circ$ vision cone. Despite the reward mechanism being designed for individual optimization, the particles' restricted sensory information results in the spontaneous emergence of collective behaviors like flocking and milling. Similarly, Grauer et al. investigate ``communicating" self-propelled particles moving in a 2D nutrient field, using DQN to optimize their sensory parameters. The trained particles exhibit three collective strategies [Fig. 7(b)]: clustering in high-nutrient regions to maximize resource consumption, spreading to minimize competition, and an adaptive strategy that dynamically switches between clustering and spreading based on nutrient distribution. Both studies demonstrate the power of RL to uncover complex self-organized behaviors without predefined interaction rules, allowing active particles to exhibit emergent collective dynamics that mimic those seen in nature.

At the macroscopic scale, RL has also been employed to regulate the self-organization in animal groups, such as fish schools and bird flocks. T. Costa et al. use evolutionary strategies, a black-box optimization method within RL, to model fish school movements, where each fish learns local interaction rules based on sensory inputs processed by a neural network. \cite{costa2020automated} By optimizing a global reward function, the agents autonomously develop four distinct collective movement patterns: rotating ball, tornado, full-core mill, and hollow-core mill behaviors [Fig. 7(c)]. Similarly, X. Wang et al. use the mean field Q-learning algorithm to model collective motion in fish schools. \cite{wang2023modeling} By representing sensory inputs as multi-channel images and designing a reward function that promotes neighbor proximity and collision avoidance, their method leads to the emergence of highly coordinated behaviors such as full-core milling and hollow-core milling, patterns often observed in nature during predation or foraging events. Moreover, M. Durve et al. apply RL to investigate flocking behaviors, where agents adjust their velocities based on the velocity information of their neighbors, resulting in the spontaneous emergence of coordinated flocking dynamics. \cite{durve2020learning} These works emphasize the role of RL in driving self-organization through individual learning. In addition, E. Nuzhin et al. apply RL to explain the emergence of swirling behaviors in animal groups, \cite{nuzhin2021animals} proposing that it arises from an escort strategy, where individuals attempt to maintain a certain distance from the group center. Their findings reveal that this self-organized strategy enhances the group's resilience to external disturbances, highlighting the survival function of collective motion.

Together, these studies demonstrate the power of RL in modeling the dynamic, adaptive nature of self-organized collective behaviors, from microscopic particles to large animal groups. By moving beyond fixed interaction rules, RL enables the exploration of complex, emergent behaviors that more accurately reflect real-world systems. Looking forward, future research could focus on refining RL algorithms to better handle high-dimensional sensory inputs and interactions in more complex environments, such as heterogeneous systems where agents have different capabilities or incomplete information. Additionally, incorporating environmental factors and constraints, such as external forces or resource availability, could further enhance the realism of these models. As RL continues to evolve, it will increasingly provide valuable insights into the mechanisms of self-organization in both natural and artificial systems.

\subsection{Goal-Directed Control of Swarm Behaviors}

Goal-directed control of swarm behaviors describes the process of deliberately directing and modulating the collective dynamics of a active swarm to achieve predefined objectives. Unlike self-organization, where behaviors emerge spontaneously through local interactions among agents without central coordination, swarm control relies on external inputs or global mechanisms to guide the system toward a specific goal. These interventions can come in various forms, such as external fields \cite{jin2021collective} and light sources, \cite{rey2023light} which influence the active particles' behaviors or the overall collective dynamics. While self-organization focuses on the emergent patterns resulting from individual-level optimization, swarm behavior control is centered on achieving targeted outcomes, such as coordinated transport or object manipulation. Regarding this aspect, RL serves as a powerful tool for optimizing control strategies by allowing agents to adapt to external inputs, overcome environmental uncertainties, and achieve complex tasks with minimal human intervention, making it an ideal approach for both controlling active swarms and using them to manipulate external objects.

Focusing first on the control of active particles themselves, M. Falk et al. employ RL to direct a swarm of self-propelled particles towards specific configurations. \cite{falk2021learning} Using a Vicsek-like model of self-propelled disks, the authors control local activity through a spotlight that increases the activity in a specific region of space [Fig. 8(a)]. They leverage actor-critic RL to train the system to achieve net transport in a particular direction. The spotlight's position and size are adjusted by the RL agent, whose decisions are based on the system's coarse-grained state, including the particle positions and velocities. The learned protocols exploit distinct physical behaviors that emerge in weak and strong coupling regimes, where the system's dynamics change as the alignment of particles is increased. Moreover, M. Schrage et al. implement RL for the directed control of ultrasound-controlled microrobots. \cite{schrage2023ultrasound} Their approach harnesses the power of ultrasound for both navigation and manipulation, utilizing the primary and secondary acoustic radiation forces to guide the microswarms. Using Q-learning, they train the system to recognize and track microrobots, adjusting acoustic forces in real-time for efficient control. As shown in Fig. 8(b), the study successfully achieves autonomous collective movements in a fluidic environment, with a specific application where the microrobots collaboratively spell out ``ETH" through ultrasound-guided motion. By using over 100,000 images, the authors teach the system to adapt to the unpredictable dynamics of the ultrasound environment, enhancing the microrobots' ability to navigate autonomously in unstructured environments.

Furthermore, to address more refined control challenges, some studies have not only focused on the precise control of active swarm itself but have also leveraged these systems to manipulate external objects. For instance, V. Heuthe et al. explore how swarm microrobots, individually controlled by laser spots, can be used to collaboratively perform complex tasks, such as rotating and transporting a large rod [Fig. 8(c)]. \cite{heuthe2024counterfactual} The RL agent is responsible for adjusting the positions of laser spots, which control the microrobots' movements, and uses a counterfactual reward mechanism within the multi-agent RL framework to assign individual credits to each microrobot. By doing so, the RL agent learns to coordinate the microrobots' actions efficiently, enabling them to interact with their environment and overcome challenges like thermal noise and complex inter-agent interactions. This approach enables the system to learn to efficiently rotate and transport objects, demonstrating the potential of active matter systems in manipulating external targets. Another related work is performed by J. Shen et al., where a deep RL environment is introduced to explore particle robot navigation and object manipulation tasks. \cite{shen2022deep} In this study, each robot is represented as a disk-shaped particle that can change its size but lacks self-locomotion capabilities. Instead of adjusting individual robots, RL algorithm optimizes the behavior of a super-agent that controls the entire swarm. The super-agent is responsible for managing the collective motion of the particle robots through switching the robot state. Using the OpenAI Gym interface, they develop a 2D simulator for various tasks, including navigation,  obstacle traversal, and object manipulation. By applying algorithms like DQN, A2C, and PPO, they benchmark the performance of particle robots in these tasks, emphasizing that these RL methods enable particle robots to interact with and manipulate external objects.

To summarize, the goal-directed control of active swarms, particularly through RL, has demonstrated considerable potential in guiding swarm behaviors and facilitating the manipulation of external objects. Despite these advancements, most of the research has been limited to 2D environments, leaving the full potential of 3D applications largely unexplored. Therefore, expanding these systems to 3D environments presents exciting opportunities and challenges. The added spatial dimension will introduce more complex dynamics in the interactions between particles and external objects, further underscoring the need for flexible and adaptive control strategies. The ability to manipulate objects in 3D environments could also open doors to more complex real-world applications, such as medical procedures, autonomous assembly, and advanced materials handling.

\section{Conclusions}

In conclusion, this review systematically explores the integration of RL in guiding and controlling active matter systems. We have discussed two key areas in this field: optimizing the motion strategies of individual active particles and regulating the collective dynamics of active swarms. For individual particles, RL has shown considerable potential in optimizing navigation, foraging, and locomotion strategies, enabling particles to adapt autonomously to dynamic environments and perform tasks such as point-to-point navigation and resource collection. At the collective level, RL has been used to regulate swarm behaviors, facilitating both self-organization and goal-directed control of swarm dynamics. In particular, RL is able to help active particles coordinate their actions to accomplish complex tasks, such as object manipulation and coordinated collective transport under challenging conditions.

While these advancements are promising, further development is still necessary to enhance the adaptability and robustness of RL algorithms in active matter systems. A key focus should be the refinement of RL algorithms to handle the complexities of active matter environments. For instance, integrating RL with multimodal sensing and feedback mechanisms will enable active particles to adapt their behaviors in response to a broader range of environmental cues, significantly expanding the potential applications of these systems. Another critical area for advancement is understanding the generalizability and transferability of RL algorithms across diverse active matter systems. Investigating how RL strategies can be effectively applied to various environments and tasks, and how well they transfer across systems, is essential for broadening their applicability. Furthermore, the development of more experimental systems of active matter that can be controlled by RL is in great demand. By addressing these challenges, we can advance the field and unlock broader applications for active matter systems, with potential breakthroughs in biophysics, robotics, medical science, environmental monitoring, and autonomous systems.

\begin{acknowledgments}

We acknowledge the support from the National Natural Science Foundation of China (Grant No. 12104147) and the Fundamental Research Funds for the Central Universities.

\end{acknowledgments}

\section*{Data Availability}

Data sharing is not applicable to this article as no new data were created or analyzed in this study.

\bibliography{aipsamp}

\end{document}